# Sanctions and Venezuelan Migration

Francisco Rodríguez[1]

Abstract: This paper examines the potential impact of different US economic sanctions policies on Venezuelan migration flows. I consider three possible departures from the current status quo in which selected oil companies are permitted to conduct transactions with Venezuela's state-owned oil sector: a return to maximum pressure, characterized by intensive use of secondary sanctions, a more limited tightening that would revoke only the current Chevron license, and a complete lifting of economic sanctions. I find that sanctions significantly influence migration patterns by disrupting oil revenues, which fund imports critical to productivity in the non-oil sector. Reimposing maximum pressure sanctions would lead to an estimated one million additional Venezuelans emigrating over the next five years compared to a baseline scenario of no economic sanctions. If the US aims to address the Venezuelan migrant crisis effectively, a policy of engagement and lifting economic sanctions appears more likely to stabilize migration flows than a return to maximum pressure strategies.

[1] Rice Family Professor of the Practice of Public and International Affairs, Josef Korbel School of International Studies, University of Denver and Director, Oil for Venezuela Foundation. E-mails: francisco.rodriguez4@du.edu, frodriguez@oilforvenezuela.org. I thank Giancarlo Bravo and Luisa García for excellent research assistance and María Eugenia Boza, Mónica García, Francisco Monaldi and Jeffrey Sachs for comments and suggestions. All errors remain my responsibility.

Contents

1.      **Introduction**

Should the United States tighten oil sanctions on Venezuela? Discussions around this issue have captured significant attention among Washington policymakers and Venezuela watchers over the past few weeks. Several opposition leaders, including Leopoldo López and María Corina Machado, have openly called for reimposing oil sanctions on Venezuela (López 2024a, 2024b; Infobae 2024), and there are some indications that current US authorities may be considering doing so (Garcia 2024; Martin 2024). During a House of Representatives hearing on December 11, 2024, Secretary of State Antony Blinken suggested that the Biden administration may be considering removing some of the licenses that it has granted for oil companies to transact with Venezuela's state-owned oil sector, stating, "Everything is on the table right now, including in terms of the licenses" (Martin 2024).[2]

This debate is a natural consequence of political developments in the South American country over the past year. On August 2, Venezuelan electoral authorities announced that Nicolás Maduro had been reelected as president for a third term by a margin of 52.0-43.2% over his opponent, Edmundo González (Consejo Nacional Electoral 2024). The country's opposition has alleged that these results are fraudulent, publishing electronic copies of tally sheets for 83.5% of polling stations providing convincing evidence that challenger Edmundo González defeated Maduro by a ratio of more than 2-to-1 (Resultados Con VZLA 2024; Kronick 2024). In the wake of what the evidence indicates was a brazenly stolen election, it is understandable that US authorities and policy analysts are discussing the best course of action.

Since 2022, the administration of Joe Biden has loosened some of the sanctions on Venezuela that were first imposed during the first term of his predecessor, current president-elect Donald Trump. This included the November 2022 issuance of a license allowing Chevron, the only US company currently active in oil production in Venezuela, to sell Venezuelan oil in the US (General License no. 41, 2022). Further loosening occurred in October 2023, when a license was issued permitting all other companies to operate under the same conditions as Chevron and expanding the scope of permitted activities to cover new investments and gas sector operations (General License no. 44, 2023). However, this license was allowed to expire in April 2024 after the US accused the Maduro government of failing to comply with the Barbados agreements on a framework for free and fair elections (US Department of State, 2024).

---

[2] As explained in section 2, the US actually grants different types of authorizations to US and non US firms to transact with Venezuela, given that it lacks jurisdiction to impede transactions between non US actors. For the purposes of exposition, we refer to these different types of authorizations as licenses, unless otherwise stated.

Because these licenses were granted in response to the resumption of talks between the Venezuelan government and the opposition, some analysts and opposition politicians have argued that the US should revoke the remaining licenses, given that the Maduro government clearly did not comply with its commitments to hold free and fair elections (López 2024b; Hernández 2024). Others, in contrast, view the US's "maximum pressure" strategy as a failed approach and argue that there is little sense in returning to a policy of isolation that has proven ineffective in changing the Venezuelan government's conduct (Banca y Negocios 2024, Rodríguez 2024a, Voght and Ventura 2024).

Among the key policy issues raised by this debate is the question of how a potential tightening of sanctions would affect other US policy goals, including those of maintaining a stable energy supply and reducing immigration flows into the United States. Over the past five years, irregular immigration into the US has risen significantly, with 11.6 million individuals apprehended by border patrol authorities while attempting to enter the country irregularly, of whom 909 thousand are Venezuelan nationals. There is a broad consensus across the US policy spectrum on the need to overhaul immigration policy, even if there are stark differences on how to achieve that goal. The incoming administration of Donald Trump has pledged to carry out a large-scale deportation program targeting irregular migrants (Hutzler 2024). However, it is unclear whether this approach can succeed without cooperation from the countries of origin to which these migrants would hypothetically be deported.

More importantly, a crucial issue in the debate is whether reimposing sanctions would exacerbate Venezuela's economic crisis, potentially increasing migration flows to Latin America and the United States. This discussion is obviously politically charged. Advocates of a hardline stance argue that Venezuelans are primarily fleeing the authoritarian regime of Nicolás Maduro and that the country's economic recovery will be impossible without fundamental political change and a return to democracy (Fuentes 2024; Kelly, Marquez and Intagliata 2024; Hernández 2024; Redacción NTN24 2024). From this perspective, the US government should maintain or even increase pressure on the Maduro regime to generate a political transition in Venezuela.

Other voices caution against the futility of this approach. They argue that the US's maximum pressure strategy towards Venezuela did not work and may instead have helped Maduro consolidate power (Voght and Ventura 2024; Ramsey and McCarry 2024; Ron 2023, Rodríguez 2024a). This aligns with empirical evidence showing that sanctions are generally ineffective at achieving regime change (Peksen and Cooper 2010; Oechslin 2014; Cohen and Weinberg 2019). In this view, increased sanctions would further deteriorate living conditions in Venezuela, driving more people to emigrate. According to this perspective, the claim that Venezuelans are fleeing solely because of Maduro's regime is nothing more than a rhetorical device that overlooks the fundamental issue: the impact of sanctions on living conditions. The more than 7 million Venezuelans who have left the country over the past decade have done so primarily because of an economic collapse that obliterated opportunities – a collapse that was caused by a combination of poor policies, mismanagement, and corruption by the regime, as well as by Venezuela's severing of financial and trading links with the rest of the world due to the imposition of economic sanctions.

This paper contributes to the debate on sanctions policy towards Venezuela by analyzing the likely effects that different sanctions policies could have on Venezuelan migration flows to Latin America, the United States, and the rest of the world. The study considers several potential scenarios of US policy changes, including the removal of the Chevron license, a return to full-fledged maximum pressure sanctions, and, on the other extreme, a complete lifting of economic sanctions. The paper estimates the impact of each of these scenarios on Venezuelan economic growth and the induced migration flows.

The findings in this paper show that economic sanctions can significantly influence Venezuelan migration patterns. Specifically, the paper estimates that a reimposition of maximum pressure sanctions would result in approximately one million additional people leaving Venezuela compared to a baseline scenario of no economic sanctions.

The remainder of the paper is organized as follows: Section 2 provides an overview of the evolution of economic sanctions on Venezuela. Section 3 discusses the evidence on the impact of these sanctions on Venezuela's economic crisis, focusing on their effects on oil production, import capacity, and productivity. Section 4 discusses the existing empirical literature on the relationship between migration and economic conditions, highlighting its relevance to the Venezuelan case. Section 5 explores various sanctions policy scenarios and outlines their potential implications for Venezuelan GDP growth and migration flows. Finally, Section 6 offers concluding remarks.

**2.     A History and Overview of US Sanctions Policy Toward Venezuela**

Venezuela's politics have been highly polarized since at least the election of Hugo Chávez to the presidency in 1998. Chávez was able to use his popularity to push through a constitutional reform in 1999 that overhauled the country's institutions, concentrating power in the executive branch. Particularly important in this respect was the ability to subordinate all other branches of government to the presidency through the threat of convening a constitutional assembly with the authority to dissolve existing branches of government. The 1999 constitution allowed Chavismo to push the opposition out of most spaces of power, consolidating a transition to a winner-take-all political system.

The concentration of political power ushered in a period of conflict over control of the country's oil revenue flows, leading to an oil strike, a major recession, and ultimately to Chávez's victory in a 2004 recall referendum. A rise in oil prices during the 2000s allowed the government to significantly expand spending and consumption and to win several elections. After Nicolás Maduro was elected to the presidency following Chávez's death in 2013, the government began to cut imports to address growing external imbalances. Oil prices nosedived in late 2014, leading the government to embark on another round of import cuts. The resulting recession allowed the opposition to capture a two-thirds majority of the National Assembly in the 2015 elections.

Emboldened by its victory, the opposition began discussing avenues to oust Maduro, including by invoking a recall referendum. After the government used its control of courts to block the referendum, the opposition decided to target the government's access to resources. In 2017, the opposition-controlled National Assembly began to publicly threaten to repudiate any debt issued by Maduro, while opposition leaders decried any attempts by international financial institutions to refinance Venezuela's debts. A campaign to denounce financing of the regime served to lobby the US government to impose sanctions barring Venezuela and its state-owned oil company PDVSA from doing business with US persons.

The first of these economic sanctions were first imposed on Venezuela in 2017, when the US government barred financing and dividend payments to Venezuela's government and state-owned oil company. Though restrictions on some state activities go back to 2005 and personal sanctions on some government officials go back to 2008, none of these were on a scale that was large or systematic enough in our view to significantly affect the functioning of the Venezuelan economy until at least early 2017.

As in the case of other countries, current Venezuela sanctions – with some exceptions – are imposed in the context of a framework created in 2015 by the Obama administration through a national emergency declaration (Executive Order 13692, 2015). However, at the time US authorities used the national emergency declaration somewhat selectively, sanctioning only seven individuals, seven of whom were security force officers allegedly involved in human rights violations around the 2014 protests (the other was a prosecutor who had filed conspiracy charges against opposition leaders; see Rodríguez 2023, p. 64).

The first step taken by the Trump administration to increase the reach of personal sanctions was to direct them at some of the highest-ranking members of the government. These began with the designation of Vice-President Tareck El Aissami in February of 2017, a month after his appointment to that post, and were subsequently increased to cover several cabinet members, supreme court justices, high-ranking PDVSA officials, the president, first lady and many of their close associates (US Department of the Treasury 2018).

On August 24, 2017, President Trump issued an Executive Order prohibiting the purchase of new debt issued by the Government of Venezuela or PDVSA and that of previously issued debt held directly or indirectly by the Venezuelan government. It also barred dividend payments to Venezuela, impeding the government from using the profits from its offshore subsidiaries to fund its budget. Exceptions were built in for short-term commercial debt, winding down of existing contracts, and transactions related to the financing of purchases of agricultural commodities or medical goods from the United States.

In November of 2018, the US issued an executive order that would lay the groundwork for more explicit trade sanctions by allowing the Secretary of the Treasury to determine that actors in a given sector of the Venezuelan economy were contributing to the national emergency. Although U.S. authorities originally presented the order as aimed at restricting trade in the country's gold sector, the order gave the government the leeway to target any sector of the economy.

Eventually, the Treasury Department would determine that four broad economic sectors were contributing to the national emergency: gold (November 2018), oil (January 2019), finance (March 2019) and defense and security (May 2019). It subsequently added several private and public sector entities belonging to these sectors to the SDN list. The designations were broad enough to essentially preclude U.S. actors from doing business with anyone in these sectors of the Venezuelan economy and thus constituted a trade embargo on almost all the country's exports. The US announced the decision to designate PDVSA in January 2019 as part of a major ratcheting up of pressure on the Venezuelan regime, with National Security Advisor John Bolton saying that he expected PDVSA to lose USD 11bn in export proceeds – a number equivalent to more than a third of the country's oil exports at the time – additional to the effect of freezing of USD 7bn in assets (De Young, Mufson and Faiola 2019). The decision was made public just five days after the US decision to recognize Juan Guaidó as the country's interim president.

Throughout 2019, the US government began to exert pressure on non-US firms to cut oil purchases from Venezuela. In August 2019, it adopted a new Executive Order that blocked any transactions with the government of Venezuela and that also gave the executive branch the power to sanction non-US persons for having "materially assisted" the Venezuelan government or its state-owned entities. From a legal standpoint, the order was essentially redundant, as all Venezuelan public sector entities at the time had their accounts blocked as a result of past sanctions or transferred to the Guaidó administration. Yet US authorities were quick to use the order to issue a stern warning to other countries, with National Security Adviser John Bolton publicly stating after the order was made public: "We want to send a message to third parties wanting to do business with the Maduro regime: There's no need to risk your business interests in the US for the purposes of profiting from a corrupt and dying government." (Kurmanaev and Jakes, 2019).

These signals, however, were insufficient to deter some relevant non-US actors from continuing to engage with Venezuela (Reuters 2019a). US authorities thus decided to ratchet up pressure on Maduro in February 2020, on the heels of Guaidó's visit to the United States. The key new decision was to sanction two subsidiaries of the Russian energy company Rosneft that had been responsible for marketing more than half of Venezuelan international sales (Mohsin and Millard 2020; US Department of Treasury 2020.) The US also sanctioned two Mexican companies that had signed oil-for-food deals with Venezuela (Kassai 2020). Rosneft at the time was handling around 75 % of Venezuela's oil sales because of other partners' caution at doing direct business with the country (Yagova, Aizhu and Párraga 2019). It had also supplied almost all the gasoline imported by the country during the previous year, as Venezuela's refining infrastructure remained beset by operational problems (Argus Media 2019).

In November 2022, OFAC issued a license allowing Chevron, the sole American firm still active in the extraction of Venezuelan oil, to produce Venezuelan oil for sale in the United States (General License no. 41, 2022). The decision was announced on the same day that the Maduro government and the main opposition coalition resumed Norway-mediated talks in Mexico City that had been suspended for over a year.

In October 2023, OFAC issued a new license, General License 44, which authorized transactions between any US persons and PDVSA (General License no. 44, 2023). In contrast to General License 41, General License 44 permitted new investments and gas sector operations. The new authorizations were extensive not only to Chevron but also to any US persons or companies. From a legal standpoint, General License 44 was equivalent to the lifting of all oil and gas sanctions on Venezuela's state-owned oil sector. However, a purely legal reading would have been insufficient to understand the scope of these authorizations. This is because US transactions with Venezuela's state-owned oil sector are limited not just by sanctions, but by the decision of the United States government not to extend formal recognition to the government of Nicolás Maduro.

In order to understand how sanctions interact with recognition, it is important to take into account that on January 23, 2019, the US announced that it would formally recognize the interim government headed by opposition leader Juan Guaidó as the legitimate government of Venezuela (Pompeo 2019). One important repercussion of this decision — which, it should be noted, implied a strong break with the diplomatic tradition of extending recognition to states and not to governments — was the transfer to Guaidó appointees of the control of Venezuela's state-owned offshore assets under US jurisdiction, as well as the right to formally represent the Venezuelan state from a legal and economic standpoint (Palladino 2019).

**Table 1: Timeline of US Sanctions on Venezuela**

| Sanction Date | Measure |
|---|---|
| March 2015 | Executive Order 13692, issued by President Barack Obama, declared national emergency |
| February 2017 | Vice-President Tareck El Aissami sanctioned; sanctions expand to high-ranking officials. |
| August 2017 | Executive Order 13808, issued by President Donald Trump, barred the purchase of new Venezuelan government and PDVSA debt and prohibited dividend payments. |
| May 2018 | Executive Order 13835, issued by President Donald Trump, prohibited US persons and entities from purchasing Venezuelan government debt, including accounts receivable, and restricted the sale or pledging of equity interests in entities owned by the Venezuelan government. |
| November 2018 | President Donald Trump issued Executive Order 13850, expanding sanctions to include the blocking of assets of individuals operating in Venezuela's gold sector. |
| January 2019 | Executive Order 13857, issued by President Donald Trump, broadened the definition of 'Government of Venezuela' to include Venezuela's state-owned oil company, PDVSA, and barred transactions with Venezuela's oil sector. PDVSA was added to the Specially Designated Nationals (SDN) List. |
| March 2019 | OFAC designates multiple entities, including financial institutions and gold sector companies, as Specially Designated Nationals (SDNs) for supporting the Maduro government. |
| August 5, 2019 | Executive Order 13884, issued by President Donald Trump, blocked all property and interests of the Government of Venezuela in US jurisdiction. It authorized secondary sanctions on entities supporting the Maduro government. |
| February 2020 | Sanctions were imposed on Russia's oil company Rosneft subsidiaries and Mexican firms for aiding Venezuelan oil sales |
| November 2022 | OFAC issued General License 41, allowing the US oil company Chevron to produce Venezuelan oil for sale in the United States |
| October 2023 | OFAC issued General License No. 44, authorizing transactions between US persons and PDVSA in the Venezuelan hydrocarbons sector. Under this license, entities conducting transactions with PDVSA are not subject to secondary sanctions. |
| April 2024 | General License 44 expired on April 18, 2024, due to Maduro's failure to fully meet commitments under the Barbados electoral roadmap agreement. A 45-day wind-down license was issued, with case-by-case review of specific license requests. |

Source: Own elaboration based on US Executive Orders, OFAC announcements and press stories.

Although the Guaidó interim government was dissolved in December 2022 (National Assembly 2022), the US continues to extend recognition for formal purposes, including the management of assets, to the opposition-controlled National Assembly elected in 2015 (Price 2023). The bank accounts and legal representation of PDVSA are also under the control of appointees of the Guaidó interim government or of the 2015 National Assembly. Because of this, even if oil sanctions were completely lifted, it would be impossible for PDVSA to carry out transactions in the United States or transactions that involve US persons, including any financial transactions through US-regulated financial institutions. One concrete implication of this is that it is not PDVSA, nor the joint ventures that it controls, that sell Venezuelan oil in the US; instead, it is Chevron or other companies that are granted, or that have been granted, licenses to do so.

It is also important to note that, from a legal standpoint, the US lacks jurisdiction to prohibit transactions between non-US actors and Venezuela's state-owned oil sector. The US, however, is able to restrict these transactions indirectly by threatening to sanction US firms that do business with non-US firms that "materially assist" the Venezuelan government or PDVSA-controlled entities. The imposition of sanctions on non-US firms for lending material assistance to Venezuela is often referred to as the imposition of secondary sanctions.

It is important to note that the US government has had the authority to impose secondary sanctions on non-US firms for doing business with Venezuela from the moment PDVSA was designated in January 2019. For transactions involving the Venezuelan government, this authority stems from Executive Order 13884 of August 5, 2019.

However, the United States has exercised this discretionary authority in varying degrees. Broadly speaking, it is possible to distinguish three different periods in terms of the imposition of secondary sanctions:

- High pressure (January 2019 – February 2020). While the US had the authority and at times issued threats of secondary sanctions during this period, the practical adoption of these sanctions was limited. In practice, their imposition was restricted to vessels that helped transport Venezuelan oil to other sanctioned destinations like Cuba.
- Maximum pressure (February 2020 – November 2022). In February 2020, the US government decided to impose secondary sanctions on two subsidiaries of the Russian oil company Rosneft for helping commercialize Venezuelan oil internationally, as well as on two Mexican companies that had signed oil-for-essentials agreements with Venezuela. This initiated a period during which the United States clearly signaled, both through these actions and informal communications, that non-US firms would likely face secondary sanctions if they were deemed to have materially assisted Venezuela. These decisions and signals led many non-US actors, such as the Indian oil company Reliance, to suspend oil trade with Venezuela. This willingness to impose secondary sanctions and to signal this stance to non-US firms continued into the first half of the administration of Joe Biden.

- Moderate pressure (November 2022- present). The US has not imposed any new secondary sanctions since the resumption of the Mexico talks and the issuance of the Chevron license in November 2022. This, coupled with announcements by several non-US firms that they have been granted licenses to operate in Venezuela's oil sector, suggests that it is appropriate to conceptualize the period starting in November 2022 as a period of loosening of US sanctions to an intensity much lower than that of the maximum pressure period.

### 3. Sanctions and Venezuela's Economic Crisis: Assessing the Evidence

Between 2012 and 2020, Venezuela's per capita GDP declined by 71%, the equivalent of three Great Depressions. This is the fifth largest contraction – and the largest peacetime contraction – documented in the cross-national data since 1950 (Table 2). It also exceeds the magnitude of all contractions documented in the Maddison Project Database between 1 and 1950 CE, making it the largest peacetime economic contraction documented in the Common Era.

In Rodríguez (2024b, 2025), I have argued that to explain Venezuela's economic collapse, one must consider the role of economic sanctions and, more broadly, political conflict. This implies going beyond the conventional narrative—which I label the policy-induced view—that attributes the collapse solely or primarily to the poor quality of the country's economic policies since Hugo Chávez first took office in 1999.

While there is significant consensus that these policies were highly distortionary and contributed to the country's low economic growth relative to its potential, it is difficult to explain the magnitude of Venezuela's economic collapse solely on the basis of poor policies for at least two reasons. The first is that the types of policies adopted by Venezuelan governments since 1999—including price and exchange controls, highly distortionary regulations, widespread nationalizations, and unsustainable fiscal and exchange rate policies—are not significantly different from those implemented by other Latin American nations in the past, including Venezuela itself.

As documented in the seminal contribution of Rudiger Dornbusch and Sebastian

Edwards (1991), this variety of macroeconomic policies often results in large recessions and fiscal and external crises, commonly associated with significant contractions in per capita income. However, the largest of these contractions, including the one experienced by Venezuela in the 1980s, typically reach around 20% to 30% of GDP—much lower than the 71% contraction observed in Venezuela during the recent collapse.

A second problem with the policy-induced view is that, as I will show later, Venezuela's productivity data reveals a steep decline occurring after 2017—precisely at the moment when political conflict in Venezuela intensified and the US imposed its first economic sanctions on the country. It is difficult to explain why policy-induced distortions stemming from policies adopted at the turn of the century would lead to such a dramatic, discontinuous drop in productivity exactly 18 years after the initial adoption of those policies.

There is a sense in which the magnitude of Venezuela's economic contraction does not appear surprising once we know what happened to oil revenues. Real oil exports fell by 93% between 2012 and 2020, by far the largest ever decline since the country began exporting oil in the 1920s. Given that oil exports accounted for almost all (96%) exports in 2012, it does not at first sight appear surprising that the near-disappearance of what was virtually the country's only oil export industry would also lead to wiping out a large share of the country's GDP.

**Table 2: GDP Per Capita Collapses, 1950-2020**

| Rank | Rank (Peacetime) | Country | Trough-to-peak ratio (percentage decline) | Period | Years | Average percentage decline | Cumulative loss (% of initial GDP per capita) | Armed Conflict |
|---|---|---|---|---|---|---|---|---|
| 1 | - | Liberia | -89.2 | 1974 - 1995 | 21.0 | -8.7 | -733.7 | Intrastate conflict |
| 2 | - | Kuwait | -86.8 | 1970 - 1991 | 21.0 | -8.1 | -1134.3 | Interstate conflict |
| 3 | - | Iraq | -77.2 | 1979 - 1991 | 12.0 | -8.2 | -365.5 | Intra and interstate conflict |
| 4 | - | D.R. of the Congo | -75.7 | 1974 - 2002 | 28.0 | -4.8 | -1190.9 | Intra and interstate conflict |
| 5 | 1 | Venezuela | -73.0 | 1977 - 2020 | 43.0 | -2.6 | -874.5 | Peacetime |
| - |  | Venezuela | -71.5 | 2012-2020 | 8.0 | -14.0 | -256.3 | Peacetime |
| 6 | - | Tajikistan | -71.4 | 1990 - 1996 | 6.0 | -18.6 | -289.9 | Intrastate conflict |
| 7 | 2 | Lebanon | -70.7 | 1974 - 1976 | 2.0 | -44.3 | -102.1 | Peacetime |
| 8 | - | Georgia | -70.6 | 1990 - 1994 | 4.0 | -25.2 | -214.8 | Intrastate conflict |
| 9 | - | Iran | -66.6 | 1969 - 1988 | 19.0 | -4.5 | -793.4 | Intra and interstate conflict |
| 10 | - | Yemen | -65.6 | 2010 - 2019 | 9.0 | -10.6 | -386.5 | Intrastate conflict |
| 11 | - | Moldova | -64.8 | 1990 - 1999 | 9.0 | -10.1 | -474.5 | Intrastate conflict |
| 12 | - | Azerbaijan | -61.0 | 1990 - 1995 | 5.0 | -16.8 | -187.5 | Intra and interstate conflict |
| 13 | 3 | Djibouti | -60.8 | 1976 - 1991 | 15.0 | -5.9 | -667.9 | Peacetime |
| 14 | 4 | Saudi Arabia | -59.9 | 1974 - 1987 | 13.0 | -6.1 | -358.9 | Peacetime |
| 15 | - | Angola | -59.2 | 1973 - 1994 | 21.0 | -4.0 | -678.2 | Intra and interstate conflict |

Note: Own calculations based on Penn World Table and own estimates of Venezuela's GDP. See Rodríguez and Imam (2022) for an explanation of the methodology to calculate growth collapses

One way of conceptualizing the role that oil has on Venezuela's economic development is through what I will call the oil well model of Venezuela's economy. According to this view — admittedly a caricature, as are all stylized economic models —Venezuela's economy I snothing more than a very large oil well. Sometimes that well produces more oil; sometimes it produces less. Sometimes that oil is worth more, sometimes it is worth less. The revenues generated by selling that oil to the rest of the world are used to make possible many other economic activities necessary to fulfill people's needs —after all, there is only so much crude oil that people can consume (not that much, it turns out). These activities range from the production of goods that are highly intensive in imported intermediates to the commercialization and sale of imported consumer goods. National accounts refer to all these activities as non-oil GDP because they add value other than by the production of crude and refined oil products. But there is a sense in which this is a misnomer, because these activities would not occur without the imported goods bought with oil revenues. We may label these sectors non-oil GDP, but 100 % of output in this economy is made possible by oil.

Economists are said to observe something working in practice and then ask whether it works in theory. This is probably an adequate characterization of how many economists would be inclined to react to the oil well model of the Venezuelan economy. While the model's explanation for the collapse appears intuitive (Venezuela lost the oil revenues that made non-oil GDP production possible), it is not easy to formalize it in a standard economics model. The reason is that in most economic models of open economies, the income of the factors involved in the production of any tradable goods is determined by the productivity of these factors and the prices that equally productive factors would receive in other countries. This is because international trade and the competition associated with it ensure that the same tradable goods do not sell for different prices in different economies that are integrated into trade.

This result, known in the trade literature as factor price equalization, has significant implications for understanding the effect of the oil sector on the rest of the economy in oil-dependent countries. If the non-oil sector employs capital, labor, and other productive factors to generate its output, then the value added by those factors will be the same as the value that equally productive factors would add in any other country. Therefore, non-oil GDP would be independent of changes in the level of oil production or the price at which oil is sold internationally.

In Rodríguez (2024b, 2025), I argue that in order to resolve this problem, it is necessary to take into account that the imports of capital and intermediate goods made possible by higher oil revenue generate what in the economics literature are known as positive externalities to the production process. That is, an additional imported intermediates or capital good will increase the production of goods in the economy that purchases them for a value that exceeds the price at which they are purchased, and market competition does not drive up the price of these imported inputs because some of the productivity gains are captured by agents different from those which purchase them — for example, because their use generates improvement in production know-how that raise aggregate level productivity. I also argue that there is significant microeconomic evidence from firm-level studies that these externalities exist and that these externalities are also recognized as being fundamental by economic growth theorists for understanding the long-term behavior of growth in per capita incomes around the world.

Once one brings in externalities associated with imported intermediates and capital goods, then one can think of the effect of oil on an economy like Venezuela's as occurring through several channels. First, oil production has a direct effect on oil GDP. If production in the oil sector declines by 80%, then by definition oil GDP at constant prices will go down by 80%. Of course, oil GDP is just a small fraction of the economy—approximately 12% over the 2010-18 period. Oil prices, in contrast, do not affect oil GDP measured at constant prices, as constant-price measures are precisely designed to abstract from any type of internally or externally generated price variations — even though they can and do affect real incomes and consumption.

Both oil prices and quantities, however, have effects on the economy's import capacity, as a higher value of exports allows the economy to import more capital and intermediate goods. Therefore, the second channel through which oil quantities and prices affect non-oil GDP is via their impact on import capacity. Lower levels of imports imply fewer externalities from these imports and a lower level of non-oil productivity.

The third channel through which oil export revenues can affect non-oil GDP is through induced accumulation of human and physical capital. If higher imports raise productivity, they will also increase the incentives to accumulate capital. Therefore, in the long run, this will lead to higher output through higher levels of investment in human and physical capital.

**Figure 1: Venezuela's per capita GDP, 1880-2023**

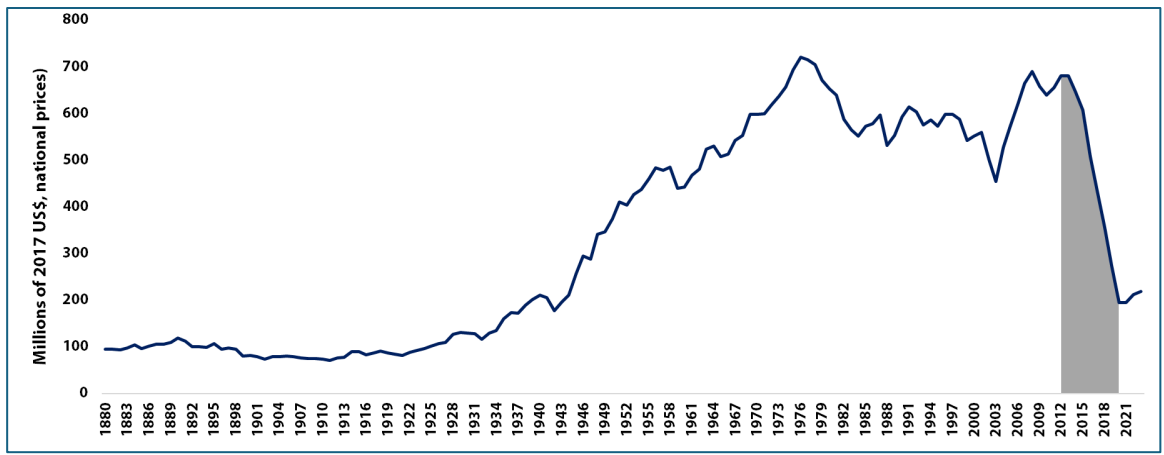

Table 3 presents the results of implementing these ideas through what is known as a standard growth accounting or, alternatively, a sources-of-growth decomposition.[3] In the conventional approach to these calculations, GDP growth is decomposed into the contributions of physical capital, human capital, and total factor productivity (TFP). These results are shown in Table 3, divided by subperiods.

We can clearly see that by far the main driver of the economy's collapse in this decomposition is the decline in productivity. GDP growth falls from 2.7% in the 1998 to 2012 pre-collapse period to -17.1% in the 2012-20 collapse period. Productivity, in turn, declines from -0.5% to -15.1% across these two subperiods. Changes in productivity thereby account for 89% of the economy's decline after 2012, according to the conventional calculation.

The last two columns of the table show the effect of including an adjustment for the externalities derived from imported inputs. These external effects explain 51.8% of the collapse after 2012, implying that, after adjusting from externalities from imports, productivity now explains a still important, but smaller, 37.1% share of the decline of output.

The last two panels of Table 3 illustrate this decomposition for a different periodization, distinguishing between the pre-sanctions period (1998-2016) and the post-sanctions period (2016-2020). What is remarkable is that, in this case, even after adjusting for imports, we still observe a productivity decline of 15.9% in the sanctions period. In fact, the decline in productivity accounts for nearly three-fifths of the decline in GDP for the post-sanctions period.

These results are also illustrated using annual data in Figure 2, which shows a clear change in trend in productivity from 2017. Adjusted for import externalities, productivity had declined steadily between 1998 and 2012 and had, in fact, risen somewhat between 2012 and 2016. After the imposition of sanctions in 2017, we see is a marked break in trend and a steady decline in import-adjusted productivity.

It is very hard to reconcile this pattern with a theory that attempts to attribute the collapse to the continued accumulation of policy distortions imposed since 1998. This inconsistency between the productivity results and the policy-induced view of the collapse becomes even starker when we consider the evidence that economic policies have become less distortionary in the post-2017 period (Batmanghelidj and Rodríguez 2021; Bull and Rosales 2020).

---

[3] Although in the text I refer to externalities from intermediates and capital goods, the calculations in Table 2 contemplate only an externality in intermediate goods. External effects from intermediate goods are assumed to affect capital accumulation through increases in the marginal product of capital , which induce greater levels of investment and a higher steady-state capital stock. While it is also possible to model direct externalities in capital goods, my empirical estimates are based on elasticity estimates derived from research on the effect of imported inputs into productivity.

**Table 3: Growth Accounting Decomposition, 1998-2020.**

|  |  | Conventional method | | | | Import external effects adjustment | |
|---|---|---|---|---|---|---|---|
|  |  | GDP | Capital | Human Capital | TFP | Imports | TFP |
| 1998-2012 (Pre-collapse) | Growth | 2.7% | 2.3% | 4.3% |  | 7.1% |  |
|  | Contribution | 2.7% | 1.3% | 1.9% | -0.5% | 2.2% | -2.6% |
|  | Percentage Contribution | 100.0% | 48.1% | 69.1% | -17.1% | 79.8% | -97.0% |
| 2012-2020 (Collapse) | Growth | -17.0% | -3.4% | 0.3% |  | -29.2% |  |
|  | Contribution | -17.0% | -2.0% | 0.1% | -15.1% | -8.8% | -6.3% |
|  | Percentage Contribution | 100.0% | 11.7% | -0.7% | 88.9% | 51.8% | 37.1% |
| 1998-2016 (Pre-sanctions) | Growth | 0.6% | 1.7% | 3.7% |  | -1.6% |  |
|  | Contribution | 0.6% | 0.9% | 1.7% | -2.0% | -0.5% | -1.6% |
|  | Percentage Contribution | 100.0% | 163.8% | 300.6% | -364.3% | -86.4% | -278.0% |
| 2016-2020 (Sanctions) | Growth | -27.0% | -6.2% | -1.4% |  | -23.1% |  |
|  | Contribution | -27.0% | -3.6% | -0.6% | -22.9% | -7.0% | -15.9% |
|  | Percentage Contribution | 100.0% | 13.2% | 2.2% | 84.6% | 25.8% | 58.9% |

Note: Calculations use Penn World Table data for 1950-2019 supplemented with own estimates for 2020.

**Figure 2: Total Factor Productivity, 1998-2020**

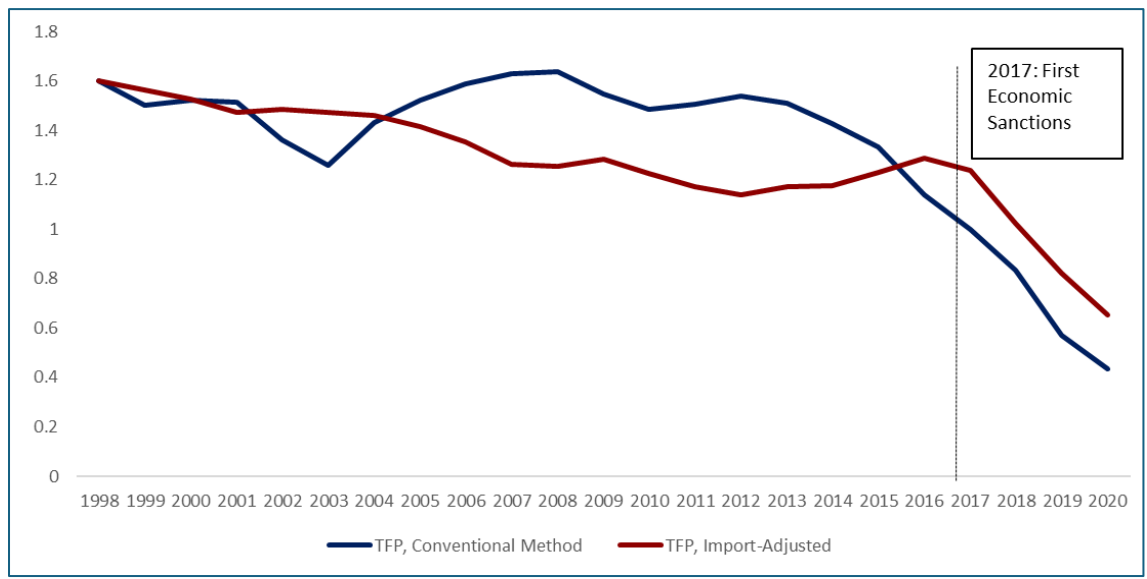

Source: Own elaboration.

I now review the evidence that 2017 and 2019 oil sanctions affected the country's capacity to produce and sell oil, thus explaining an important part of the reduction in oil GDP and import capacity. As discussed above, the decline in Venezuela's oil revenues by 93% is the most important proximate cause of the country's growth collapse in the 2012 to 2020 period. It thus makes sense to begin our analysis of sanctions by focusing on their effect on oil revenues. Figure 3 plots the evolution of Venezuela's oil production between 2008 and 2020, according to data reported by secondary sources to OPEC. This series shows remarkable stability up to 2016. Production begins falling in 2016, with the rate of decline accelerating markedly over the subsequent four years. Venezuela's production level at the end of 2020, at around 400 tbd, was about one-sixth of its pre-2015 production.

The series represented in Figure 3 also suggests that there are three distinct periods in the evolution of Venezuela's oil production data. Production remains relatively stable at around 2.3 million barrels up to December 2015. At the start of 2016, it begins to decline and falls at a rate of 1.0% per month. Then, from September 2017 on, the rate of decline accelerates, averaging 3.1% over the following sixteen months. After the imposition of oil sanctions, it suffers two discrete jumps: a 35.2% drop (405 tbd) between January and March 2019 and a 55.7% drop (423 tbd) between February and June 2020. The series sees another discontinuous break in trend from November 2022, with production now beginning to increase after a period of stability in previous months.

**Figure 3: Oil production, 2008-2020**

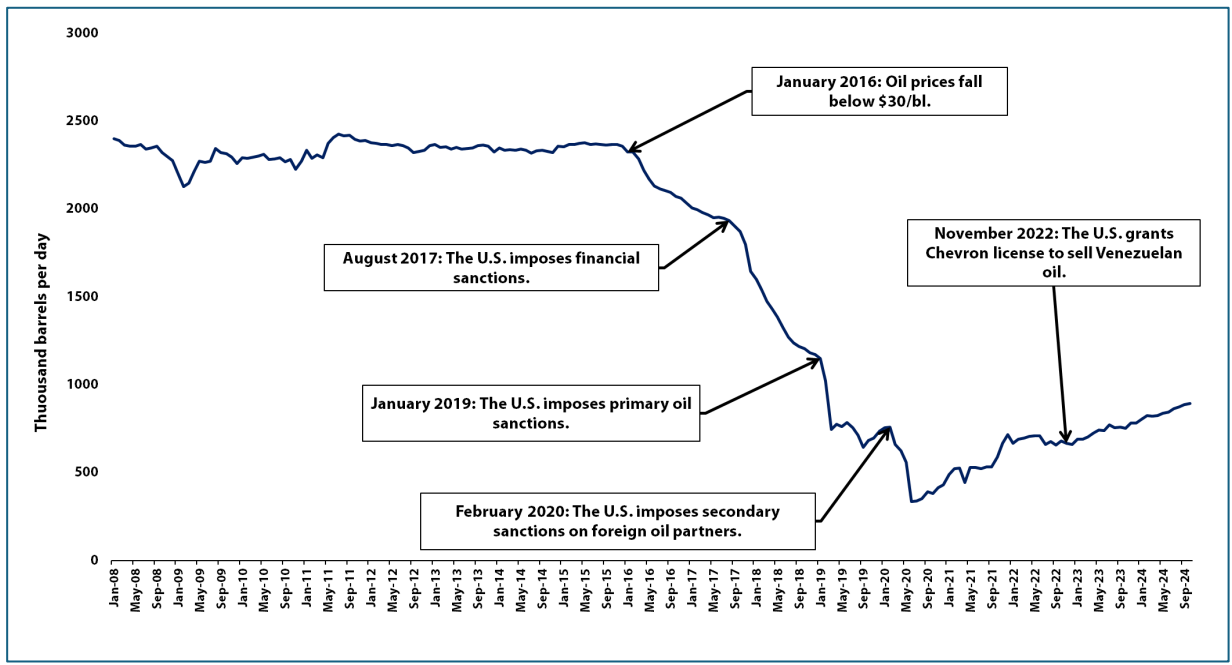

Source: OPEC Secondary Sources.

Some of the key inflection points in this figure follow episodes of imposition of new sanctions or tightening of existing ones. The first acceleration of the rate of decline occurs after the adoption of financial sanctions restricting the Venezuelan government and the state-owned oil company Petróleos de Venezuela (PDVSA) from issuing new debt, refinancing existing debt, or receiving dividends from US subsidiaries. The second inflexion point occurs right after the US Treasury Department's Office of Foreign Assets Control (OFAC) designated PDVSA, effectively banning any oil trade with Venezuela. The third inflexion point occurs shortly after the US imposes secondary sanctions on subsidiaries of the Russian company Rosneft, which at the time was handling 75% of Venezuela's international oil sales) (Yagova, Aizhu, and Parraga 2019). The fourth inflexion point – which, in contrast to the previous ones, marks an increase instead of a decrease in the rate of growth of oil production — occurs after the adoption of General License 41 to allow Chevron to sell Venezuelan oil in the US. While the first inflexion point in early 2016 does not correspond to a sanctions event, it does coincide with the slump in oil prices that saw the price of a barrel of Venezuelan oil fall to less than $25 (from $97 in the first half of 2014) and is similar in magnitude to the declines seen by other high-cost oil producers during the same period (Rodríguez 2019).

Rodríguez (2024b, 2025) discusses the results of several econometric estimates of the effects of sanctions on Venezuela's oil production (readers interested in the technical details of these estimations should refer to those publications). Three different types of empirical estimates are considered: synthetic control methods, cross-country panel data estimates, and firm-level difference-in-differences estimators.

The synthetic control method constructs a counterfactual scenario to estimate what Venezuela's oil production would have been in the absence of sanctions. This method involves selecting a weighted combination of other oil-producing countries whose determinants of production trends closely match Venezuela's before the imposition of financial sanctions in August 2017. The synthetic group serves as a benchmark or "synthetic Venezuela" to approximate the trajectory that Venezuela's oil production would have followed if sanctions had not been imposed.

The analysis finds that by December 2018, Venezuela's actual oil production had fallen 53 log points below the synthetic counterfactual. This represents a loss of approximately 797,000 barrels per day, translating to a revenue loss of $17.5 billion at an average oil price of $60 per barrel. The results demonstrate a clear and substantial deviation from the synthetic control group after sanctions were implemented, suggesting that sanctions played a critical role in the collapse of oil production.

A second approach uses a cross-country two-way fixed effects panel regression to compare the performance of 39 sanctioned and non-sanctioned oil-exporting countries over the 1999-2020 period. The results indicate that sanctions are associated with a decline of between 69 and 73 log points in oil production. For Venezuela, this corresponds to a reduction of approximately 572,000 to 594,000 barrels per day, equivalent to a revenue loss of around $16

billion.

The third approach uses firm-level panel data to analyze the determinants of monthly production data from 33 production blocs operating in Venezuela's Orinoco Belt region. Using a difference-in-differences (DiD) approach, the production trajectories of firms that had access to finance are compared with that of those that did not. Pre-sanctions financial access is measured through the existence of pre-sanctions Special Financing Vehicles (SFVs) agreements in which joint-venture partners lent to state-controlled firms that they were minority shareholders in. This identification strategy is premised on the observation that SFVs should have been disproportionately affected by the sanctions, which cut off access to international financial markets and thus prohibited SFV agreements. The results show that firms with SFVs experienced sharper production declines after the imposition of sanctions. Specifically, the analysis estimates that the typical Venezuelan joint venture firm suffered a production loss of 39,000 to 51,000 barrels per day per firm, representing a 48-54% decline relative to pre-sanctions production levels.

In Rodríguez (2024b), I summarize the effects of these and other estimates of the impact of sanctions on oil production. The range of estimates is broad, ranging from 29 to 104 % of the decline in production attributable to sanctions. In order to model the effects that this has on the broader economy, I adopt an intermediate level corresponding to the midpoint between the average level of two varieties of calculations of these impacts, yielding an estimated contribution of sanctions that equals 50.3% of the observed decline in production.

I then introduce these estimates into an economic model calibrated to the productivity decompositions presented in Table 3. The results are presented in Table 4 below. The sanctions-induced decline in oil output affects GDP through two channels. First, since oil accounted for 12% of GDP prior to the collapse, there is a direct contribution of sanctions to the GDP decline through reduced oil output. This represents a loss of 4.1 percentage points of initial (2012) GDP. Second, there is an indirect effect through the losses in non-oil GDP caused by reduced import capacity. This explains 33.0 percentage points of the GDP decline. Of this, around three-fifths is due to lower production, with the other two-fifths due to price declines, implying that an additional 8.7 percentage points of the loss of GDP can be attributed to reduced import capacity caused by sanctions.

The politically-induced loss of access is incorporated by considering three sources of funding that the country would have surely made use of had it been able to. These are the Special Drawing Rights issued by the IMF in its 2021 general allocation, which were valued at $5.0 billion; the IMF's non-program Rapid Financing Instrument facility, which would have given the country

access to 150% of its IMF quota, or $7.8 billion; and its Central Bank gold and other reserve assets blocked at the Bank of England, valued at $2.1 billion. Assuming that in 2020 the country would have spent one-third of these assets in higher imports, the politically induced loss of access to external funding sources is found to account for a loss of 2.8 percentage points of GDP.

Last, I consider the effects of sanctions on total factor productivity. Using the difference between the post-2016 decline in productivity (15.9% annual) and the 1999 to 2016 decline in productivity (1.6%) to estimate the share of productivity in the 2017 to 2020 period attributable to sanctions and other politically induced toxification effects delivers an estimate of 18.8 percentage points of 2012 GDP, with another 5.6 percent attributable to the indirect effects through factor accumulation.

The sum of all these effects is 40.0% of 2012 GDP, or 55.8% of the decline in output observed in the 2012 to 2020 period. Of this loss, 61% is attributable to direct and indirect productivity effects, 32 percent to sanctions effects on oil production, and 7 percent to loss of access to external funding sources.

This numerical approximation suggests that more than half of the decline in GDP observed in Venezuela between 2012 and 2020 can be attributed to politically induced causes, including economic sanctions, the loss of access to external funding sources, and the politically induced toxification of relations with the Venezuelan economy. One way to think about this number is by noting that it implies that in the absence of these economic effects of political conflict, Venezuela's economy would have contracted by 31.6% in the 2012 to 2020 period. Such a contraction would have been more in line with the magnitude of other large contractions in developing countries, as well as in past Venezuelan history, that were induced by a combination of external shocks with prior unsustainable macroeconomic policies.

In other words, if we ask how much of Venezuela's economic contraction between 2012 and 2020 can be attributed to sanctions and other politically-induced restrictions on the country's insertion into global markets, the results in this section suggest that the answer is around half. But if we ask why Venezuela suffered such an unusually large collapse, a good case can be made that sanctions are the primary cause. Without the severing of the country's trade and financial links with the global economy, Venezuela would still have undergone a major economic crisis after the collapse in oil prices. But not only would that crisis have been much smaller than what we observed; it would also not have been atypical given the magnitude of the terms of trade decline and the country's exposure to external shocks. Had Venezuela's income declined by around a third after seeing a fall of more than two-thirds in oil prices, not many observers would have considered the result unusual. The fact that Venezuela suffered an economic collapse of a much greater order of magnitude is what makes the country's experience atypical.

**Table 4: Decomposition of 2012–2020 economic growth in politically induced and other drivers.**

| Concept | Cumulative percent decline, 2012-2020 | Percentage contribution |
|---|---|---|
| **Change in per capita GDP** | **-71.5%** | **100.0%** |
| **Oil GDP** | **-8.1%** | **11.3%** |
| Sanctions effect | -4.1% | 5.7% |
| Other causes | -4.1% | 5.7% |
| **Non-oil GDP** | **-63.4%** | **88.7%** |
| Import capacity | -32.9% | 45.9% |
| Oil exports | -30.0% | 42.0% |
| Oil price | -12.7% | 17.8% |
| Oil production | -17.3% | 24.2% |
| Sanctions effect | -8.7% | 12.2% |
| Other causes | -8.6% | 12.0% |
| Permanent loss of access to credit | -2.8% | 3.9% |
| TFP | -23.5% | 32.9% |
| Sanctions and toxification effects | -18.7% | 26.2% |
| Other causes | -4.8% | 6.7% |
| Factor Accumulation | -7.0% | 9.8% |
| Sanctions and toxification effects | -5.6% | 7.8% |
| Other causes | -1.4% | 2.0% |
| **Aggregates** | | |
| **Change in per capita GDP** | **-71.5%** | **100.0%** |
| **Sanctions and toxification effects** | **-39.9%** | **55.8%** |
| **Other causes** | **-31.6%** | **44.2%** |

Note: Table decomposes Venezuela's 2012-2020 contraction into the contribution of sanctions and other politically induced restrictions on the country's insertion into global markets.

## 4. Growth, Development and Migration: A Review of the Literature

I now turn to discussing how Venezuela's unprecedented economic collapse affected its emigration flows to the rest of the world. It may seem intuitive to expect that a large collapse in GDP would drive mass emigration flows. In fact, Venezuela's emigration data strongly supports the idea that the economic collapse is the primary—and perhaps even the only relevant—driver of the country's mass exodus.

Figure 4 shows three different emigration series for Venezuela from 2010 to 2024. While there are some important differences among the series, which are discussed in greater detail in the appendix, they all tell a broadly similar story. Venezuelan emigration to the rest of the world was relatively small in the years prior to 2015, with existing series coinciding in placing it well below 100,000 persons per year. Emigration began to accelerate markedly in 2016, the country's first year of double-digit negative economic growth. In that year, the number of people leaving the country rose to around 300,000.

The economy would then suffer four additional consecutive years of double-digit negative growth. In 2017, emigration flows rose to around 800,000 people, and then increased markedly to well above one million in 2018. The series, discussed in greater detail in the appendix, began to diverge more noticeably at this point. The emigrant stock data shows a massive 2.2 million people leaving in 2018, while other series report a still large but more restrained 1.4 million. The emigrant stock data, which is generated based on a combination of flow and stock statistics from recipient countries, tends to show systematically larger emigration flows for 2018, 2019, and 2020. However, all three series agree that the period from 2016 to 2021 represents an era of unprecedentedly high emigration rates, unlike anything seen in Venezuela's past.

The series also differ in terms of the timing of the slowdown. The emigrant stock data shows sustained high levels of emigration in 2022 and 2023, while the other flow series indicate levels consistent with pre-crisis emigration trends. Nevertheless, by the end of 2024, all series agree that Venezuela has seen emigration rates return to pre-crisis levels.

The fact that this normalization of emigration rates occurred following four consecutive years of positive economic growth—coinciding with increasing oil production and the easing of sanctions—strongly supports the view that economic factors, including sanctions, are primary drivers of Venezuelan emigration.

**Figure 4: Emigration from Venezuela and per capita GDP, 2010-2024**

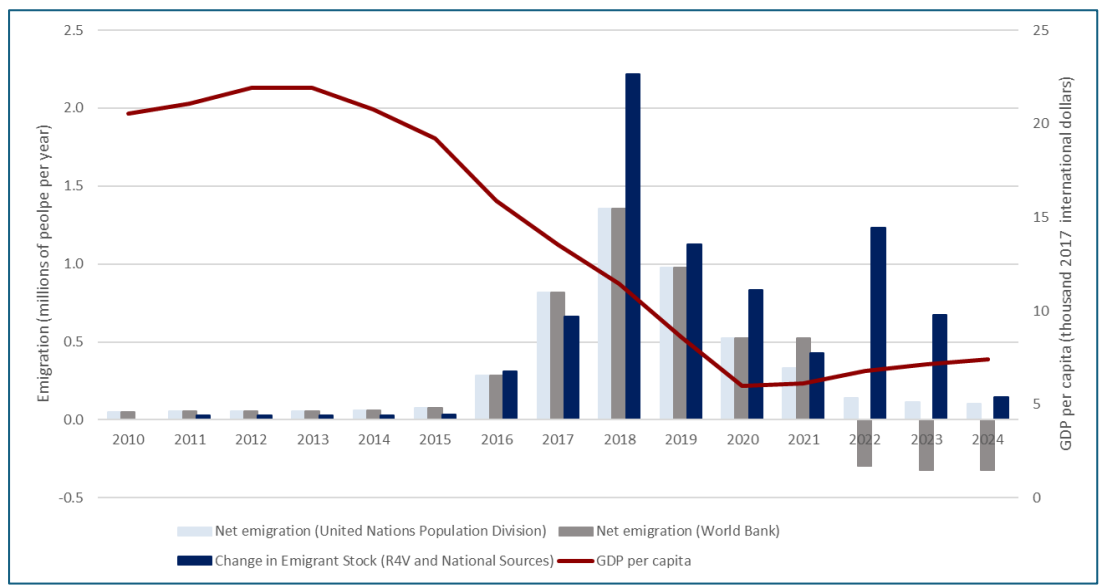

Source: Own elaboration.

How does this story fit into the economics literature on the relationship between incomes and migration? The issue is more complex than one might think at first sight. Certainly, a long-standing tradition in economics, going back to the seminal work of Todaro (1969), posits that migration is primarily driven by economic disparities between places of origin and destination. According to this view, poverty, unemployment, and lack of opportunity are the root causes of migration, and development policies that improve economic conditions in origin countries should reduce out-migration (Massey et al. 1993; Borjas 1989). This "root-causes" approach underpins many policy initiatives by advanced country governments, which aim to mitigate migration pressures through job creation, poverty alleviation, and infrastructure development (National Security Council 2021).

However, a number of scholars have pointed to an important stylized fact that is problematic for the root causes approach – the tendency for emigration rates to be highest at intermediate levels of development. This phenomenon, known as the "migration hump" hypothesis, shows that there is an inverted-U relation between development and migration in the cross-sectional data, with emigration being an increasing function of development levels at low levels and only beginning to decline as countries surpass a certain threshold of income. This pattern is consistent with microeconomic evidence that shows that the poorest individuals tend to have lower propensities to emigrate, partly due to the fact that migration entails a significant economic investment. As countries move from low to middle-income status, some argue, improved economic conditions provide individuals with the means and aspirations to migrate (Martin and Taylor 1996; Clemens 2014). De Haas (2010) further emphasizes that development expands people's capabilities, making migration a more viable option in the short-to-medium term.

Whether development reduces outmigration, as suggested by the traditional root-causes approach, or whether it can in some cases lead to greater emigration flows, as predicted by the migration hump hypothesis, continues to be an area of controversy. In a recent paper, Bencek and Schneiderhienze (2024) argue that the migration hump result is an artifact of the data driven by systematic differences between poor and middle-income countries in unobserved determinants of migration. They show that using a panel data fixed effects specification — which controls for time-invariant country-level characteristics — the migration hump result disappears, and a negative, statistically significant, relationship between per capita GDP and emigration to OECD countries emerges in the data.[4] Clemens and Mendola (2024), in contrast, have used survey data for nationally representative samples in 99 developing countries to estimate the effect of individual income on the propensity to emigrate, as proxied by whether respondents claim to be actively preparing to emigrate. They find that there is a significant positive relationship between respondents' income and the propensity to emigrate, consistent with the hypothesis that poverty may be an impediment to emigration.

To understand the implications of this discussion for the case of Venezuela, it is important to understand that the migration hump hypothesis is concerned primarily with the long-term effects of development on migration, where development is understood in a broad sense to refer not just to income levels but also to all of the other structural changes that take place in a society that sees significant expansions of capabilities. Advocates of the migration hump hypothesis believe that, as a poor country undergoes the structural changes that are associated with sustainable increases in per capita income, these changes will lead to increases both in the capabilities and in the aspirations for emigration of people in these countries. Therefore, it is important to understand that the migration hump hypothesis is (i) a theory about the long-term effects of development on emigration, which need not be the same as short-term effects; and (ii) is a theory about the effects of economic and social development broadly conceived instead of the partial effects of income, which is an imperfect proxy for one dimension of development.

The last point can perhaps be best illustrated by a key result of Clemens and Mendola (2024), which is the finding of a negative relationship between the propensity to migrate and income within education groups. This result, illustrated in Figure 5 below, shows that better-off individuals are less likely to prepare for emigration than lower-income individuals with the same level of education. Despite this, it is still the fact that higher-income individuals are on average more likely to emigrate, because higher-income individuals are more likely to have higher levels of education, and people with higher levels of education are more likely to emigrate than those with lower levels of education amd the same level of income.

Consider the implications of the findings in Figure 4 on what we could expect to happen with emigration in the case of Venezuela. If Venezuela were a country that were gradually and steadily advancing in its development – as it did throughout much of the 20$^{th}$ century, then we would expect not only its incomes to grow but its population over time to become more educated, and thus, according to these findings, more propense to emigrate. Similarly, if Venezuela underwent a sustained, long-term trend of impoverishment, then it may be the case that younger cohorts who grow up in a poorer country than their parents will have less possibilities to acquire education and will ultimately be less likely to emigrate.

---

[4] In response to an earlier version of this paper, Clemens (2020) had argued that the results could be driven by spurious relation between the regressors, yet Bencek and Schneiderhienze (2024) show, using Monte Carlo simulation methods, that the statistical bias in practice from the nonstationarity of variables has little impact on the coefficient estimate.

But these long-term effects are distinct from what we may expect to happen if Venezuela suffers a serious economic shock, such as a collapse in oil exports – as occurred during the past twelve years. In that case, the country's average level of education will not change – as it is determined by the past choices and conditions of those who are alive today. Therefore, both the less-educated and more-educated groups will move to the left along the lines represented in Figure 5, leading to an increase in emigration, It is true that if Venezuela never recovers from this economic decline, its population may eventually end up being less propense to emigrate – but that is a change that will take at least one generation to fully materialize.

**Figure 5: Emigration demand during structural transformation**

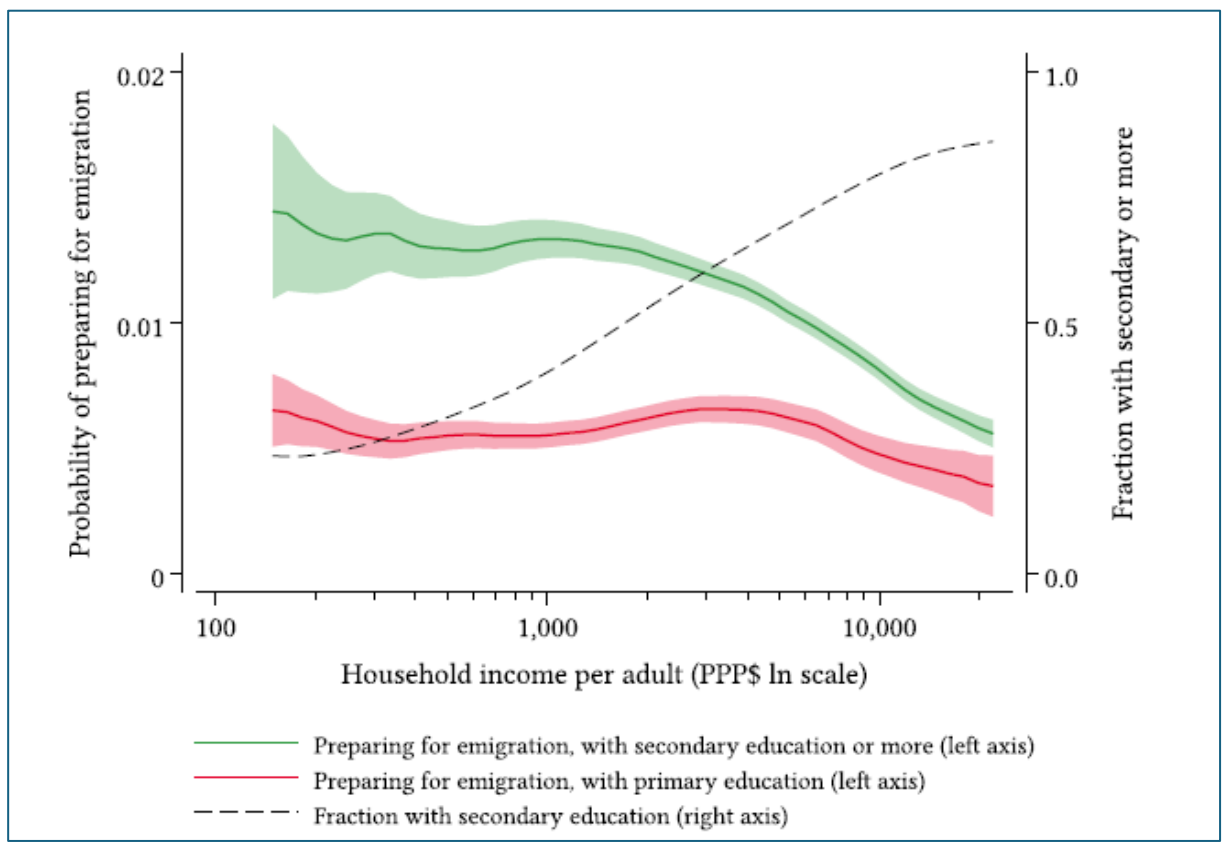

Source: Reproduced from Clemens and Mendola (2024).

This example also helps illustrate why there is not necessarily a contradiction between the predictions of migration hump theory – that, for countries below a certain threshold level of income, development raises emigration and economic decline can lower emigration – and the fact that emigration tends to rise during many economic crises, both in poor countries and in advanced economies. Figure 6 shows how emigration rates and per capita incomes have evolved under four large crises: Venezuela (2010-2021), Zimbabwe (1995-2017), Greece (2005-2020) and Mexico (1980-1994[5]).

**Figure 6: Emigration rates and per capita incomes, selected cases**

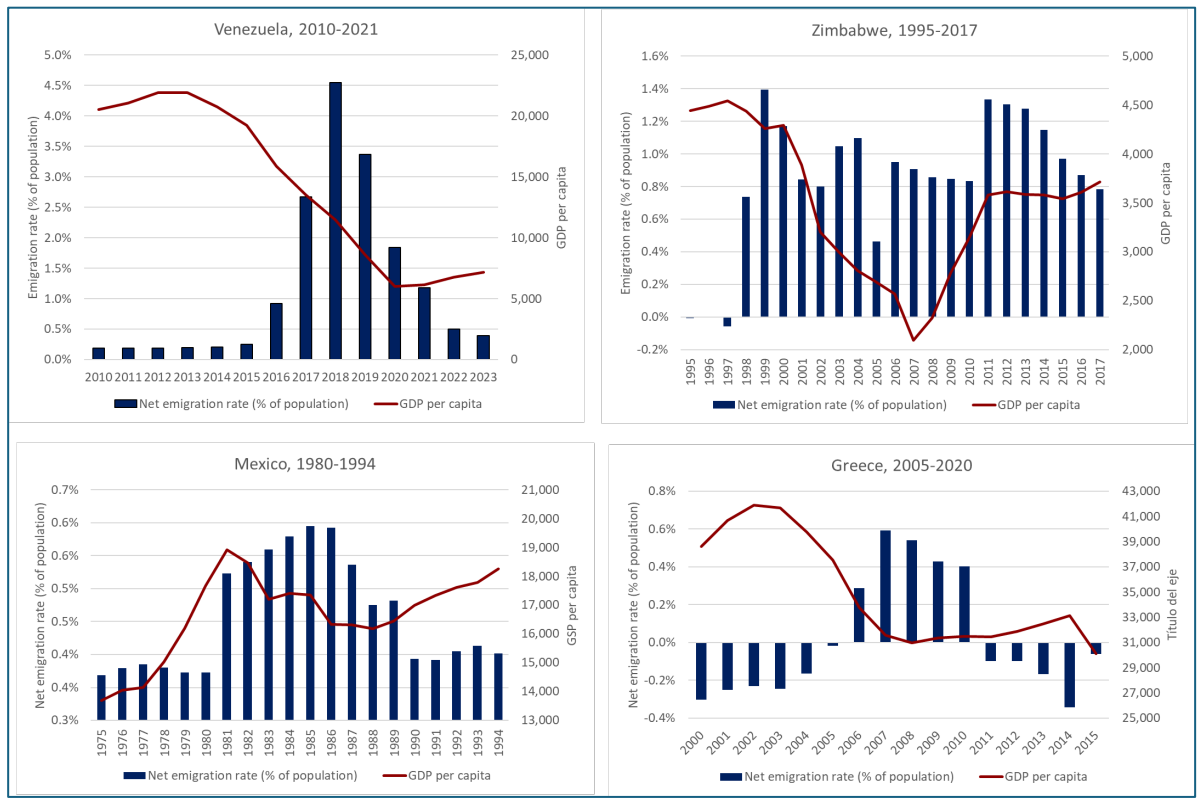

Sources: Own elaboration based on World Bank and UN Population Division data.

The migration and income data shown for the four case studies in Figure 5 illustrate how economic crises, characterized by sharp declines in GDP per capita, are often accompanied by significant increases in emigration rates. In the case of Venezuela, net emigration rates remained low until a few years into the economic collapse. As GDP began its steep decline, emigration surged, peaking in 2018. Emigration rates remained high in 2019 and 2020 before declining to much lower levels as the economy began to stabilize in 2021-2023. Similarly, Zimbabwe saw historically low or even negative net emigration rates before the beginning of its economic downturn in the late 1990s. The subsequent collapse in GDP triggered a significant rise in emigration, which remained elevated even after the economy recovered post-2008, suggesting the presence of structural factors sustaining outflows. For Greece, the financial crisis led to a shift from net immigration to net emigration, with migration rates returning to pre-crisis patterns as the economy recovered. Finally, Mexico experienced a surge in emigration during the debt crisis of the 1980s, followed by a return to baseline migration levels once economic stability was restored.

---

[5] For venezuela, we use the average emigration rate across the three series shown in Figure 4.

There is a sense in which the increases in migration for some of these cases are not necessarily contradictory even with the long-term predictions of the migration hump hypothesis – although they are, of course, also consistent with a more traditional root causes approach. According to the migration hump hypothesis, for countries above a certain threshold level of income — estimated by Clemens (2020) to be around $10,000 in 2011 PPP-adjusted prices — declines in per capita GDP are associated with increases in emigration. As shown in Figure 7, both Greece and Mexico fit this framework well; during their respective crises, their per capita incomes were above this threshold, and the resulting economic downturns spurred substantial increases in emigration. In contrast, Zimbabwe, which was below this threshold during its economic collapse, could not be that easily explained in the migration hump hypothesis.

**Figure 7: GDP per capita (PPP 2011$) compared to migration inflection threshold, selected cases**

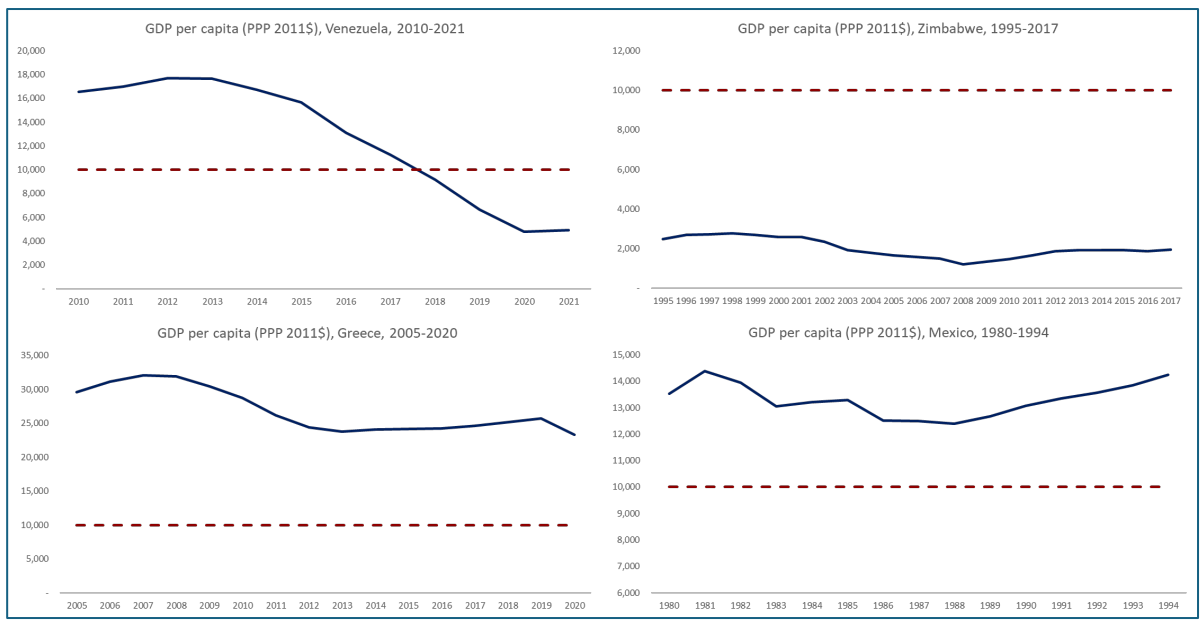

Sources: Own elaboration based on World Bank, Penn World Tables 10.01, and Clemens (2020).

Regarding Venezuela, at the onset of its economic crisis in 2012, its per capita income ($17,678 2011 PPP-adjusted US dollars per person) was well above the threshold, meaning that the subsequent economic collapse would have been expected to drive up emigration rates.

Indeed, Venezuela's net emigration surged significantly between 2015 and 2017 as GDP per capita plummeted. By 2018, Venezuela's income level crossed below the $10,000 threshold. While the migration hump hypothesis would suggest that further economic declines after that moment would lead to lower emigration rates, this interpretation applies only in a short-term, literal sense. As discussed above, the migration hump hypothesis primarily explains long-run development trends rather than immediate responses to economic shocks. In Venezuela's case, the magnitude and speed of the economic collapse from 2012 to 2020 occurred while the country was still at a development level — in the broad sense, which goes well beyond short-term income levels — where significant emigration was predicted. Therefore, the observed emigration patterns are consistent with both the migration hump hypothesis and the more traditional root-causes approach, where economic decline spurs migration.

I now consider the short-term relation between emigration and GDP implicit in existing panel data. To do so, I use a panel of data on emigration rates, per capita incomes and population for the 1960-2019 period. I estimate the following regression specification:

$$e_{it} = \alpha_0 + \alpha_1 g_{it} + \eta_i + \epsilon_{it} \qquad (1)$$

where $e_{it}$ represents net emigration as a percentage of the population for country $i$ at time $t$, $g_{it}$ is the logarithmic growth of GDP per capita for country $i$ at time $t$, $\eta_i$ captures a country-specific fixed effect, and $\epsilon_{it}$ is the idiosyncratic disturbance term. The coefficient estimates for this equation are shown in Table 5. If the regression is estimated on the whole sample, we get a coefficient of -1.0%, which indicates that a ten percent change in GDP is associated with an increase of .1 percent in the emigration rate of the country of origin. For a country like Venezuela, with an emigration rate of 0.6 percent of the population in the most recent years (2021-2023), this would imply an increase if around one-sixth in the emigration rate.

However, the coefficient estimate on the whole sample may not be adequate for capturing the case of a country that is undergoing a large economic crisis, as has been the case of Venezuela. Table 5 also shows that the coefficient estimate increases to 2.2% when we restrict the sample to countries experiencing negative GDP growth, and rises further to 2.9% for cases of countries experiencing intense economic crises – defined as yearly drops in GDP by more than 5%. In the last row of the table I present a coefficient estimate based only on the Venezuelan data for the 2011-2019 period.[6] That coefficient, which captures the increase in emigration that would occur if Venezuela's emigration reacted to changes in GDP similarly to how it did over the past decade, is 5.2%, indicating that a 10 percent decline in GDP could lead to a near-doubling of emigration rates.

---

[6] This is calculated as the ratio of the changes in emigration rates to the log changes in GDP between the 2011-2013 and 2017-2019 periods.

**Table 5: Effect of a one percentage increase in GDP on net emigration rates**

| Sample | Coefficient |
|---|---|
| Complete sample | -0.010% |
| All crisis episodes | -0.022% |
| Large crisis episodes | -0.029% |
| Venezuelan crisis episode | -0.052% |

Source: Results from estimation of equation (1) on cross-national panel data. See appendix for estimation details.

## 5. Analysis of Policy Scenarios

In this section, I analyze how different policy scenarios regarding US economic sanctions towards Venezuela could impact migration flows. To achieve this, I combine the analysis of the effects of sanctions on economic growth presented in section 3 with the regression estimates of the effects of economic crises presented in section 4. Integrating these two strands of analysis, I estimate how various US policy decisions — including maintaining, lifting, or intensifying sanctions — might influence Venezuelan migration flows. This approach allows us to understand the potential consequences of sanctions on migration patterns, providing a framework for assessing the broader human and economic implications of US economic statecraft.

I start by briefly summarizing the analysis presented in section 3. Various estimation methods were used to assess how sanctions have constrained oil production, drawing on techniques such as synthetic control methods and difference-in-differences estimation. The findings indicate that sanctions directly reduced oil output, a critical driver of Venezuela's economy, which in turn contributed to the country's significant GDP contraction. The impact was not limited to the oil sector; sanctions also affected non-oil GDP by reducing the country's import capacity, thereby constraining access to essential intermediate goods.

Drawing on those results, I now simulate the effects of various US sanction scenarios on the Venezuelan economy and migration flows. I first estimate the implications of each sanction scenario for oil revenues by analyzing the expected changes in oil production and oil prices — with the latter effect derived from the observed effect that sanctions have had on the discount at which Venezuelan oil sells in international markets.

Next, I estimate how changes in oil revenues affect the country's export levels and, consequently, its capacity to finance imports. Given the importance of imported intermediate goods to Venezuela's non-oil sector, an increase in imports can generate productivity gains through the external effects discussed in section 3.

Additionally, I consider the impact of restoring access to international credit markets. I assume that Venezuela would gain access to approximately $3 billion annually in lending or access to previously frozen assets (e.g., gold reserves or SDRs) if sanctions are lifted – one-fifth of the current estimated stock of external assets not subject to legal attachment procedures. This increased import capacity would further boost productivity in the non-oil sector, reinforcing the positive spillover effects on the broader economy.

If all economic sanctions are lifted, I also assume that the country would recover the lost productivity associated with sanctions and discussed in section 3. This assumption is based on the view that most of these negative productivity effects stem from the economy's incapacity to engage in international transactions. In that case, there is an alternative channel through which the lifting of sanctions would affect the economy above and beyond growth in output or import capacity, which is the direct recovery in non-oil productivity.

Table 6 below outlines our four distinct policy scenarios related to US sanctions on Venezuela and their corresponding impacts on oil production, oil prices, and export levels. These scenarios help us estimate how potential changes in sanctions policy might influence Venezuela's import capacity and subsequently its GDP and migration flows.

- Scenario 1: Status Quo

This scenario assumes a continuation of the current policy environment, particularly the existing framework under General License 41 and specific licenses granted to other companies. Using as a baseline recent production numbers, I estimate that oil production growth into 2025 will continue its current growth trajectory. I estimate a price of oil of $68 per barrel in this scenario, which is consistent with recent levels and trends.

- Scenario 2: Removal of General License 41:

I simulate an intermediate scenario in which he US government revokes General License 41, which currently permits Chevron to operate in Venezuela. In that case, I estimate a decline of 174,000 barrels per day — equivalent to the increase in the Chevron joint ventures production that came after the approval of General License 41, net of revenue flows that go to debt repayment.[7] This would reduce total production to approximately 775,000 barrels per day. I assume that the discount on Venezuelan oil would rise to its level in the period immediately preceding the issuance of General License 41, reducing the sales price of Venezuelan oil to $65 per barrel.

- Scenario 3: Return to Maximum Pressure:

In this scenario, the US reinstates maximum pressure sanctions similar to those applied during 2020-2021. I assume that oil production reverts to the levels observed in 2021, when production was significantly constrained by sanctions, resulting in an output of around 550,000 barrels per day. The 2021 period is used as a reference for production levels – in order not to penalize it with production levels which also fell due to the COVID pandemic. However, the 2020 discount on Venezuelan oil of 2020 is used in this scenario, as that discount is likely to better reflect the effect of sanctions on the willingness of buyers to acquire Venezuelan oil. This scenario thus assumes a sales price of $54 for a barrel of Venezuelan oil.

- Scenario 4: Lifting of All Economic Sanctions:

Drawing on the findings discussed in section 3, I estimate that lifting all economic sanctions would restore production to 1,663,000 barrels per day. In this scenario, there are two additional effects of sanctions easing on the economy: first, the economy regains access to capital markets and offshore assets and can thus increase imports by more than the increase in oil revenues; second, the economy recovers productivity losses associated with sanctions. It is important to note that the scenario is intended to capture a full, comprehensive and sustained lifting of all policy restrictions on trade with Venezuela or the Venezuelan government. Therefore, it includes both the lifting of financial sanctions, and the recognition of the Venezuelan government by international financial institutions such as the International Monetary Fund. This scenario, while possibly not realistic in strict political terms, is still extremely important in order to provide us with a counterfactual that can serve and calculate the total effect of sanctions compared to a scenario.

**Table 6: Oil production, prices, discounts and exports under different scenarios**

---

[7] Oil production in Chevron joint ventures rose from 50 thousand barrels per day prior to the issuance of GL 41 to 240 thousand barrels per day in the most recent data. I estimate that of these revenues, that corresponding to some 80 thousand is currently being used for debt repayment. Since Chevron's debt is projected to be repaid by the end of 2025 (Párraga and Buitrago, 2023), this implies that the difference between the production that generates increases in import capacity in scenarios 1 and 2 would over the five-year span used in these projections would be of approximately 16 thousand barrels per day, yielding our estimate of 174 thousand barrels.

| Scenario | Dates | Discount | Production | Oil price | Oil exports |
|---|---|---|---|---|---|
| Status quo | May 2024 – November 2024 | 20.6% | 949.4 | 67.9 | 19,083,353 |
| Removal of GL 41 | February 2021- October 2022 | 24.6% | 775.4 | 64.6 | 14,029,960 |
| Return to maximum pressure | March 2020- December 2020 | 36.8% | 551.0 | 54.1 | 7,325,853 |
| Lifting of all economic sanctions | January 2013-July 2017 | 12.9% | 1,663.2 | 74.5 | 40,359,337 |

Source: Own elaboration based on scenarios discussed in text.

Table 7 presents the results of our simulations under different US sanctions scenarios and their corresponding effects on imports, GDP, and migration flows. Increased oil exports directly translate into higher imports of goods and services, as shown in the imports column of the table. These increases in imports, in turn, positively affect GDP due to their impact on non-oil productivity through externalities associated with intermediate goods.

In order to estimate the immigration consequences of different sanctions regimes, I use the coefficient estimates presented in Table 5 above. As shown there, the choice of sample can have a significant effect on the magnitude of the reaction of emigration rates to declines in per capita GDP. I present three alternative calculations: the first calculation uses the coefficient estimate of the equation for all economic crises; the second one for large economic crises in which GDP is declining at more than 5% a year; and the third one uses the coefficient derived from the increases in immigration associated with Venezuela's economic decline during the past decade. The last column presents the average of those three calculations. I also use as a baseline the GDP level estimate for 2025 published by the IMF in its *World Economic Outlook*.

**Table 7: Economic and migration impacts of sanctions scenarios**

| Scenario | Imports of goods and services | GDP Growth (2025-2029) | Emigration (2025-2029) | | | |
|---|---|---|---|---|---|---|
| | | | Conservative | Intermediate | Historical | Average |
| Status quo | 19,638,907 | 2.0% | 816,328 | 816,328 | 816,328 | 816,328 |
| Removal of GL 41 | 14,585,513 | 0.4% | 866,469 | 883,585 | 945,896 | 898,650 |
| Return to maximum pressure | 7,881,406 | -2.0% | 939,241 | 981,199 | 1,133,947 | 1,018,129 |
| Lifting of all economic sanctions | 43,893,901 | 17.8% | 329,937 | 163,903 | -440,549 | 17,764 |

Source: Own elaboration based on scenarios discussed in text.

The results of each of the scenarios are summarized in Tables 7 and Figure 8. Some key highlights are:

- Scenario 1 (status quo): Under the continuation of current policy, the Venezuelan economy would grow at 2.0 % a year. In that case, we expect current migration trends to continue, with approximately 160 thousand persons emigrating each year, for a cumulative emigration of 0.82 million persons in the next five years (2025-2029).

- Scenario 2 (removal of General License 41): In this scenario, oil production would decline by 174 thousand barrels per day relative to baseline. This would generate a decline of around 8 percent of GDP relative to the baseline, lowering the growth rate over the next five years by 1.6 percent a year. This would generate additional outflows of around 80 thousand additional emigrants (50-130 thousand depending on scenarios)

- Scenario 3 (return to maximum pressure). This scenario generates a much steeper fall in GDP, causing a decline of 19% relative to baseline – implying a growth rate of -2.0% over the 2025-29 period, four annual percentage points below the baseline. In this case, the number of Venezuelans who emigrate would rise to 0.94-1.13 million people over the next five years, or an increase of 123-318 thousand persons over the current baseline.

- Lifting of all economic sanctions. In this scenario, oil production recovers one-half of the ground lost since before sanctions, rising to 1.66 million barrels per day by 2029. The economy also recovers the productivity losses attributable to sanctions in the post-2017 period, as well as access to capital markets, allowing it to restructure its debts and use some of its external assets. In this scenario, imports would more than double in a five-year period, and GDP would rise by a factor of 2.2 – implying growth of 17.8% a year. In this case, migration flows would decline significantly. In the more conservative scenario, they would fall to 330 thousand emigrants – around two-fifths their current levels. In the historical scenario, based on the sensitivity of migration to incomes seen in Venezuela's recent experience, Venezuela would experience return migration of 88 thousand persons a year. A simple average of the three estimates yields emigration rates of 18 thousand emigrants, or approximately 2 percent of that in the baseline scenario.

**Figure 8: Migration impact of different sanctions scenarios.**

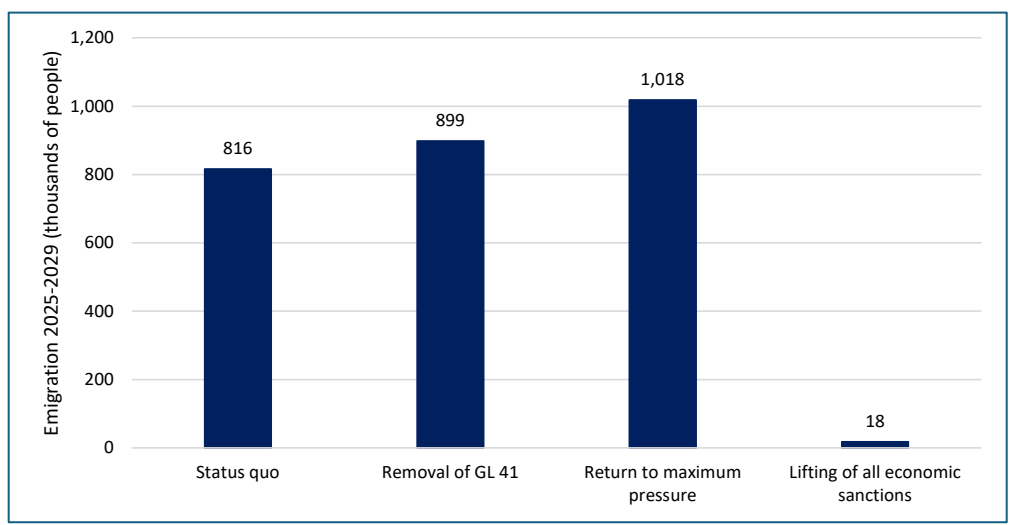

Source: Own elaboration based on results reported in Table 7.

One way to summarize the findings of these scenarios is by comparing the results under the third and fourth scenarios, which represent the most extreme stances of policy: a return to maximum pressure versus a lifting of all economic sanctions. This comparison provides a reasonable estimate of the migration effects of maximum pressure strategy compared to a counterfactual of sanctions. The difference between these two scenarios is a migrant flow of one million Venezuelan emigrants over the next 5 years. That is this paper's estimate of the migration impact of maximum-pressure sanctions.

## 6. Concluding Comments

This paper has discussed the potential effects on Venezuelan migration of changes to US economic sanctions currently under consideration. It has presented several policy scenarios and employed various estimation methodologies to approximate these effects. The central conclusion of this analysis is that US sanctions policy significantly impacts Venezuelan migration. Sanctions damage the Venezuelan oil sector, which generates the vast majority of the country's foreign currency revenue. Consequently, sanctions affect the economy not only directly, through their impact on the oil sector, but also indirectly by reducing import capacity, which limits the availability of essential imported intermediates and capital goods for the non-oil sector.

The evidence presented decisively shows that sanctions have been a major contributor to Venezuela's economic collapse, with politically motivated foreign policy decisions explaining approximately half of the decline in living standards observed since 2012. Our estimates suggest that a tightening of economic sanctions would lead to significant increases in the number of people leaving Venezuela over the next five years. Specifically, policies such as the removal of the Chevron license or a return to maximum pressure would have discernible effects on migration.

It is also the case that these effects would be small compared to the massive migration flows witnessed over the past decade – I particular when one considers deviations from the current baseline policy. This is partly a consequence of Venezuela's current status as a heavily sanctioned country: even a return to maximum pressure would not dramatically alter the economy's limited access to international markets. Therefore, the most meaningful comparison is not between the current status quo and the removal of the Chevron license or a return to maximum pressure, but between these scenarios and the complete lifting of economic sanctions. Our analysis indicates that a lifting of economic sanctions would essentially erase Venezuelan emigration flows, and could well even generate some return migration. In contrast, a maximum pressure strategy would result in approximately one million additional Venezuelan migrants leaving the country over the next five years compared to a scenario where all economic sanctions are lifted.

The implications of these findings for US foreign and domestic policy are profound. If the goal of the US government is to mitigate the Venezuelan migrant crisis and political change in Venezuela is assumed unlikely, the most effective policy would be to lift all economic sanctions. Beyond their effect on migration, there are other compelling reasons to lift sanctions. As discussed here and in related research, economic sanctions have significantly contributed to Venezuela's economic collapse and, consequently, the suffering of its people. Additionally, there is no evidence that sanctions have successfully pressured the Maduro regime; if anything, they appear to have strengthened its grip on power. This outcome aligns with extensive empirical research indicating that sanctions are largely ineffective at driving regime change, while engagement tends to offer better prospects for fostering long-term democratic transformation (Cohen and Weinberg 2019; Peksen and Cooper 2010; Oechslin 2014).

Furthermore, the US stands alone among major economies in imposing economic sanctions on Venezuela. The European Union, for instance, has consistently refused to implement economic sanctions, arguing that such measures would exacerbate the humanitarian crisis (European Parliament 2023). This divergence underscores the challenge of pursuing a multilateral sanctions strategy. If the US aims to coordinate its policy with other significant global actors, it should reconsider its unilateral approach to sanctioning the Venezuelan economy.

Some have argued, and will continue to argue, that one should not assume political change in Venezuela is impossible and that sanctions should be maintained because they are necessary to bring about such change. The flawed reasoning in this argument, however, is that there is nothing in the existing literature or in Venezuela's experience to suggest that sanctions positively contribute to the prospects for democratization. In fact, the evidence indicates that many of the key drivers of democratization—both globally and in Venezuela—are internal rather than external (Teorell 2010, Lucan and Way 2012, Pop-Eleches and Robertson 2015; Wejnert 2014). These factors include the opposition's ability to coordinate effective strategies, the willingness of moderates on both sides of the political spectrum to propose and work toward viable political solutions, and the presence of internal fissures within the regime.

Surely, it is possible that a combination of these internal factors could generate political change in Venezuela. In such a scenario, it is highly likely that Venezuela's economy would recover significantly as a result of both the lifting of sanctions and the implementation of improved economic policies and social reforms. What is flawed from a positive standpoint, however, is the assumption that the likelihood of political change depends on maintaining economic sanctions.

More importantly, from a normative standpoint, it is deeply problematic to believe that it is justifiable on moral grounds to inflict significant and lasting damage on the lives of vulnerable Venezuelans under the premise that the increased suffering caused by sanctions is justified by the end goal of political change. This reasoning disregards the ethical implications of policies that deliberately worsen living conditions for millions in the pursuit of uncertain political outcomes.

el aspirante Edmundo González Urrutia, 5" *X* , August 2, 1:46 p.m. https://x.com/cneesvzla/status/1819459661114102120.

De Haas, Hein. 2010. "The Internal Dynamics of Migration Processes: A Theoretical Inquiry." *Journal of Ethnic and Migration Studies* 36 (10): 1587–1617.

De Young, Karen, Steven Mufson and Anthony Faoila. 2019. "Trump Administration Announces Sanctions Targeting Venezuela's Oil Industry." *The Washington Post*, January 28. https://www.washingtonpost.com/national/health-science/trump-administration-announces-sanctions-targeting-venezuelas-oil-industry/2019/01/28/4f4470c2-233a-11e9-90cd-dedb0c92dc17_story.html

Dornbusch, Rudiger and Sebastian Edwards. 1991. *The Macroeconomics of Populism in Latin America.* University of Chicago Press.

European Parliament. 2023. "Impact of sanctions on the humanitarian situation in Syria." https://www.europarl.europa.eu/RegData/etudes/BRIE/2023/749765/EPRS_BRI(2023)749765_EN.pdf

Executive Order 13692. Blocking Property and Suspending Entry of Certain Persons Contributing to the Situation in Venezuela. 2015. 80, CFR 47. 2015. https://www.govinfo.gov/content/pkg/FR-2015-03-11/pdf/2015-05677.pdf

Fuentes, Fernando. 2024. "María Corina Machado advierte sobre una "ola de migración sin precedentes" si Maduro se aferra al poder." *La Tercera*, August 8. https://www.latercera.com/mundo/noticia/maria-corina-machado-advierte-sobre-una-ola-de-migracion-sin-precedentes-si-maduro-se-aferra-al-poder/M6IYFHRNJVHKTLAR3MJF3IAZWU/

Garcia, David. 2024. "US considering new visa curbs, oil sanctions on Venezuela amid post-election standoff." *Reuters*, September 24. https://www.reuters.com/world/us-studying-new-visa-restrictions-oil-sanctions-venezuela-amid-post-election-2024-09-27/

General License No. 41: Authorizing Certain Transactions Related to Chevron Corporation's Joint Ventures in Venezuela. 31 CFR part 59. Department of the Treasury. 2022. https://ofac.treasury.gov/media/929531/download?inline
38

____. 2020. "Venezuela sanctions: Frequently ask questions."
https://ofac.treasury.gov/faqs/817#:~:text=On%20February%2018%2C%202020%2C%20OFAC,sector%20of%20the%20Venezuelan%20economy.

Voght, David and Patricia Ventura. 2024. "There's a more effective way forward than "maximum pressure" for Venezuela." *Atlantic Council*, December 3. https://www.atlanticcouncil.org/blogs/energysource/theres-a-more-effective-way-forward-than-maximum-pressure-for-venezuela/

Yagova, Olga, Chen Aizhu and Marianna Parraga. 2019. "Rosneft Becomes Top Venezuelan Oil Trader, Helping Offset U.S. Pressure." *Reuters*, August 22.
https://www.reuters.com/article/russia-venezuela-oil-idUKL8N24Y4Z6

Wejnert, Barbara. 2014. "Factors Contributing to Democratization." In *Diffusion of Democracy: The Past and Future of Global Democracy*, edited by Barbara Wejnert. Cambridge University Press, 2014.

World Bank. 2024. World Development Indicators.
https://databank.worldbank.org/source/world-development-indicators
43

## 8. Appendix on Data Sources

I use three sources to estimate Venezuelan emigration. The first is the series on net migration published by the World Bank in its World Development Indicators (World Bank 2024). The second is total net migration published by the United Nations Population Division (United Nations 2024a). While the World Bank series is based on the UN Population Division series, they differ substantially in some cases, including Venezuela's, as shown in Figure 4. This appears to be due to modeling assumptions, including smoothing decisions made by the World Bank.

The third data source is a combination of existing databases on migrant stocks of Venezuelan immigrants in destination countries. This is based on a series for 17 Latin American countries published by the Interagency Coordination Platform for Refugees and Migrants from Venezuela (R4V 2024). I complement this with data on US foreign-born Venezuelan nationals from the American Community Survey (Bureau of the Census 2024) and data on Venezuelan-born nationals in Spain from the Spanish Statistical Agency (INE 2024). Between 2010 and 2015, I use migrant stock data from the UN Migrant Stock Database for all countries (United Nations 2024b). From 2017 onward, I use the R4V data for the 17 countries covered by that database, US data from the American Community Survey, and Spanish data from the Spanish Statistical Agency. Data for all other countries is presumed to grow after 2015 at a rate proportionate to the ratio between their 2010–2015 growth rate and that of other countries in the UN Migrant Stock Database. I use data from the December version of the R4V database when available. When not available, the number is interpolated from the most closely adjacent months. Estimates for 2000 and 16 are interpolated between the 2015 UN series and the 2017 series estimated through the methods just outlined.

Data for cross-national regressions comes from the World Population Prospects and net migration series, World Development Indicators GDP per capita at constant 2017 prices (PPP-adjusted), and the population series from the World Development Indicators database. In the case of Venezuela, where the World Development Indicators database does not have GDP data after 2014, I use the Penn World Table series, updated with growth rates after 2019 from the International Monetary Fund's World Economic Outlook Database (IMF 2024).